\documentclass[12pt]{article}
\input{psfig}

\newcommand{\BABARPubYear}    {00}

\newcommand{\BABARProcNumber} {32}
\newcommand{\SLACPubNumber} {8780}

\input babarsym

\begin{document}
{\pagestyle{empty}

\begin{flushright}
SLAC-PUB-\SLACPubNumber \\
\babar-PROC-\BABARPubYear/\BABARProcNumber \\
February, 2001 \\
\end{flushright}

\par\vskip 4cm

\begin{center}
\Large \bf Studies of CP Violation at \Lbabar\ 
\end{center}
\bigskip

\begin{center}
\large 
{Anders Ryd~\footnote{ryd@hep.caltech.edu}} \\
{\it California Institute of Technology, 356-48, Pasadena CA 91125} \\
(for the \lbabar\ Collaboration)
\end{center}
\bigskip \bigskip

\begin{center}
\large \bf Abstract
\end{center}
BABAR has studied the time dependent asymmetries in the the decays
$B^0\to J/\psi K_S^0$ and $B^0\to \psi(2S) K_S^0$ in a data set
of 9.0 fb$^{-1}$ taken at the $\Upsilon(4S)$ resonance.
In these channels 
we reconstruct 168 events of which 120 are flavor tagged and used in
a likelihood fit where we determine $\sin2\beta$. The flavor of the 
other neutral $B$ mesons is tagged using information primarily from
identified leptons and Kaons. A neural network is used to recover
events without any clear Kaon or lepton signature. 
A preliminary result of $\sin2\beta=0.12\pm0.37\pm0.09$ is obtained.

\vfill
\begin{center}
Contributed to the Proceedings of the \\
7$^{th}$ International 
Conference on $B$-Physics at Hadron Machines, \\
9/13/2000---9/18/2000, Sea of Galilee, Isreal
\end{center}

\vspace{1.0cm}
\begin{center}
{\em Stanford Linear Accelerator Center, Stanford University, 
Stanford, CA 94309} \\ \vspace{0.1cm}\hrule\vspace{0.1cm}
Work supported in part by Department of Energy contract DE-AC03-76SF00515.
\end{center}
 
}

\newpage

\section{Introduction}

 One of the main goals of the BABAR experiment is to study CP Violation
in neutral $B$-mesons. The neutral $B$ meson system is similar to the
neutral Kaon system in that we have two flavor eigenstates 
that mix. However, the phenomenology is rather different. In the Kaon
system the physics is driven by a large difference in the decay widths, such
that there are two states, the $K_L^0$ and the $K_S^0$, that have 
substantially different lifetimes.
In the $B_d$ system the widths of the two states are very similar and
instead the physics is dominated by the mass difference, $\Delta m_{B_d}$,
which controls the oscillation frequency of the $B^0\bar B^0$ system. 

 In the Standard Model, CP Violation arises from complex phases in the
CKM matrix. By studying the interference between decays of $B$ mesons
that decays directly to a common final state and those that mix 
before they decay, we can study the phases of the CKM 
triangle\cite{babarphysbook}.

\begin{figure}
\centerline{\psfig{figure=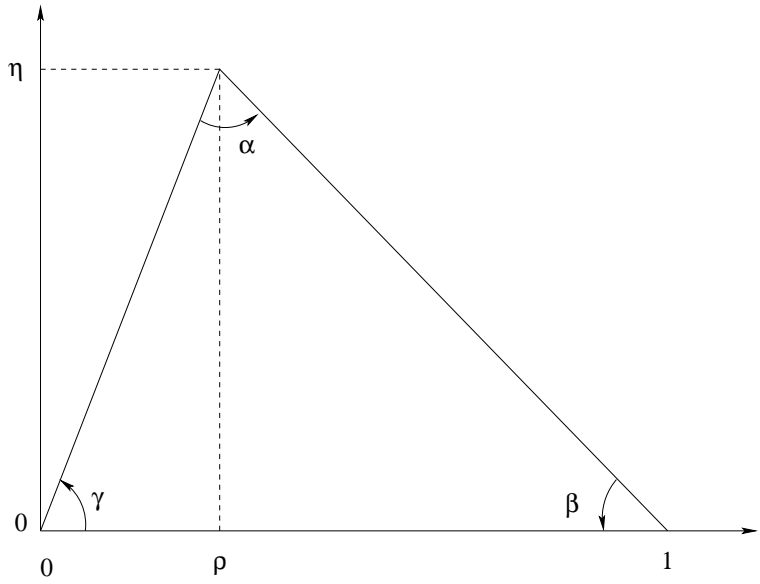,height=3.0in}}
\caption[]{
The CKM triangle formed by unitarity between the first and
third generation. The study of time dependent asymmetry in 
$B^0\to J/\psi K_S^0$ allows us to determine $\sin2\beta$.
}
\label{fig:ckm}
\end{figure}

 Unitarity of the CKM matrix allows the construction of unitarity 
triangles. Applying the unitarity constraint between the first and
third generation gives the least degenerate triangle. The angles
of this unitarity triangle 
are probed in CP Violation in $B$ decays. Figure~\ref{fig:ckm} shows
the normalized CKM triangle, the studies of CP Violation in 
$B\to J/\psi K^0_S$ allow us to determine $\sin2\beta$.

 The decays of primary interest in this presentation are 
$B^0\to J/\psi K_S^0$ and the very similar $B^0\to \psi(2S) K_S^0$.
These decays have been dubbed the golden modes for measuring $\sin2\beta$.
These modes are experimentally fairly straight forward to construct,
and have branching fractions that allow us to collect samples of
events that are sufficiently large to study the time dependent asymmetry.
But foremost, they are the golden modes due to the very small theoretical
uncertainties as there are no penguin contributions with different
weak phases.

 The time dependent rate for $B\to J/\psi K^0_S$ is given by
\begin{equation}
f_{\pm}(\Delta t;\Gamma,\Delta m_{B_d},\sin2\beta)=
       {1\over 4}\Gamma e^{-\Gamma |\Delta t|}
       [1\pm \sin2\beta\times \sin\Delta m_{B_d}\Delta t],
\label{eq:sin2betaorig}
\end{equation}
where $+(-)$ indicates that the $B_{\rm tag}$, the other $B$ from the
$\Upsilon(4S)$ decay, was tagged as
a $B^0$ $(\bar B^0)$. To experimentally fit the time 
distribution we need to account for two effects; finite
resolution in the $\Delta t$ determination and the possibility that
the wrong tag was assigned to $B_{\rm tag}$. The time resolution
is handled by convoluting Eq.~\ref{eq:sin2betaorig} with a 
resolution function, ${\cal R}(\Delta t;\hat a)$. The fraction of
mistags, $w$, dilutes the $\sin2\beta$ measurement by 
${\cal D}=1-2w$. Experimentally we will perform a fit to
\begin{equation}
{\cal F}_{\pm}(\Delta t;\Gamma,\Delta m_{B_d},{\cal D}\sin2\beta,\hat a)=
       f_{\pm}(\Delta t;\Gamma,\Delta m_{B_d},{\cal D}\sin2\beta)
       \otimes {\cal R}(\Delta t;\hat a).
\label{eq:sintwobeta}
\end{equation}
  
The time dependent CP asymmetry 
for the $B^0\to J/\psi K_S^0$ and
$B^0\to \psi(2S) K_S^0$ decays is given by
\begin{equation}
a_{CP}(\Delta t)={N_{B^0}(\Delta t)-N_{\bar B^0}(\Delta t) 
             \over N_{B^0}(\Delta t)+N_{\bar B^0}(\Delta t) }
={\cal D}\sin2\beta\sin(\Delta m_{B_d}\Delta t),
\end{equation}
where $N_{B^0}$ ($N_{\bar B^0}$) is the number of events where $B_{\rm tag}$
was assigned a $B^0$ ($\bar B^0$) tag.
In these expressions $\Delta t=t_{CP}-t_{\rm tag}$ refers to the 
time difference between the decay of $B_{CP}$ and $B_{tag}$. 
At the $\Upsilon(4S)$ where the 
two $B$ mesons are produced in a coherent state, the flavor of the $B$
decaying to the $CP$ eigenstate is determined by studying the flavor
of the other $B$, as the two $B$ mesons are in a coherent $P$-wave
state we know that the flavor of $B_{CP}$ is of the opposite 
flavor of the $B_{\rm tag}$ at the time of the $B_{\rm tag}$ decay.
Hence, $\Delta t$ is the time $B_{CP}$ evolved from when we knew
its flavor. 
Further,
at PEP-II the collisions are asymmetric, this allows us to measure the 
difference between the decay times of the two $B$ mesons
by simply measuring the separation
in $z$ of the two $B$ decays, 
$\Delta t \approx \Delta z/c\langle \beta\gamma \rangle$.

 The major steps in performing this analysis are
\begin{itemize}
\item Reconstruct the CP eigenstates.
\item Measure the vertex resolution, ${\cal R}(\Delta t;\hat a)$.
\item Determine vertex separation, which gives $\Delta t$
\item Tag the flavor of the other $B$, $B_{\rm tag}$.
\item Measure the wrong tag fraction, $w$.
\item Perform likelihood fit to the $\Delta t$ distribution to determine 
      $\sin2\beta$.
\end{itemize}
As far as possible we try to determine the resolution function
parameters and 
wrong tag fractions from data. 

\section{The BABAR experiment}

 The BABAR experiment is located at the PEP-II storage ring at SLAC. 
PEP-II collides electrons and positrons with energies of about 9.0 GeV/c$^2$
and 3.1 GeV/c$^2$ respectively at the center of mass energy of the
$\Upsilon(4S)$ resonance. The produced $\Upsilon(4S)$ mesons have a 
boost of about $\beta\gamma=0.56$. Since the first recorded collisions
with the BABAR experiment on May 26, 1999, PEP-II has 
produced excellent luminosity that have allowed BABAR to collect the 
worlds largest sample of $B$ mesons
at the $\Upsilon(4S)$. The analysis presented
here is based on 9.0 fb$^{-1}$ recorded on the resonance and 0.8 fb$^{-1}$
taken about 40 MeV/c$^2$ below the resonance. This corresponds to about
$10.5\times 10^6$ produced $B\bar B$ pairs. (The 2000 run of BABAR ended
on October 30, 2000, and the total recoded luminosity in 2000 was 23 fb$^{-1}$.)

 The BABAR experiment is described in detail elsewhere~\cite{babarpaper}. Here
just a few key points of particular relevance to the measurement 
presented will be discussed.

\begin{figure}
\centerline{\psfig{figure=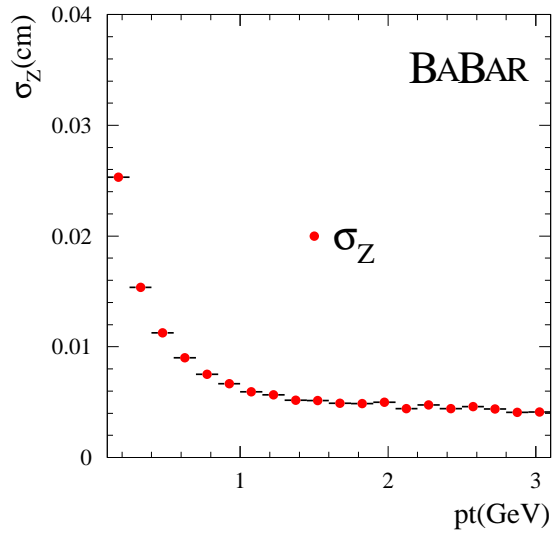,height=3.0in}}
\caption[]{
The impact parameter resolution in $z$ for the silicon vertex tracker as
measured by cosmic muons.
}
\label{fig:svtres}
\end{figure}

 One of the key features of this experiment is that the produced $B$
mesons have a boost of about $\beta\gamma=0.56$, and that the time difference
between the $B$ decays is measured by the separation in $z$ position
of the $B$ decay vertices. 
The typical vertex separation between two $B$ meson decays is 250 $\mu$m.
The BABAR experiment has a 5 layered, 
double sided, silicon micro strip vertex detector capable of
stand-alone tracking for low momentum particles, $p_T<120$ MeV/c, 
not detected in the drift chamber. 
Figure~\ref{fig:svtres} shows the measured track impact parameter
resolution in $z$ as a function of momentum. At higher momenta, where 
multiple scattering is negligible, the resolution is about 40 $\mu$m.
For an exclusively reconstructed $B$ the vertex resolution in $z$
is typically about 40-60 $\mu$m.

\begin{figure}
\centerline{\psfig{figure=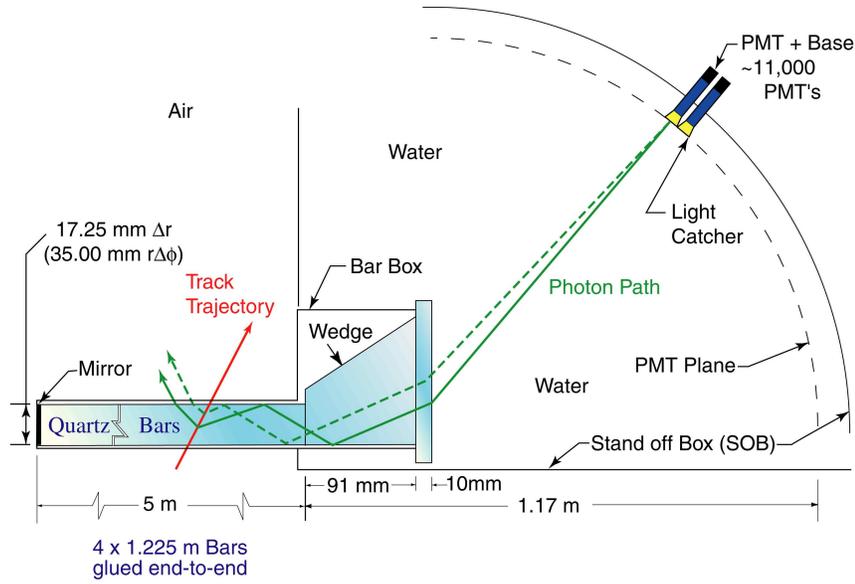,height=3.0in}}
\caption[]{
The principle of the DIRC; incident charged particles emits Cherenkov
radiation as they pass through the quarts bars. The Cherenkov photons
are reflected to the end of the quartz bar and out in a water tank,
which is instrumented with photomultipliers to detect the light.
}
\label{fig:dirc}
\end{figure}

 Another unique feature of the BABAR experiment is the charged
hadron identification system, the DIRC, Detection of Internally Reflected
Cherenkov light. The DIRC detects Cherenkov
photons that are produced in quartz bars and reflected out to a 
water tank instrumented with photo multipliers, see Figure~\ref{fig:dirc}.
The DIRC has proved to work very well, we have achieved better than 3$\sigma$
$K-\pi$ separation at  momenta of 3 GeV/c. Further improvements are
possible with a better understanding of the alignment of the
DIRC with respect to the tracking system.

 Overall the BABAR experiment has performed very well, important for our
ability to record data is our efficiency, typically BABAR is
live 97\% of the time when PEP delivers luminosity. Losses are due to
background spikes that cause trips, ramp-up of voltages, and sporadic 
outages,
e.g., of computing resources. After the data is recorded, 
it is typically reconstructed within 24 hours.

\subsection{Reconstruction of event samples}

\begin{figure}
\centerline{\psfig{figure=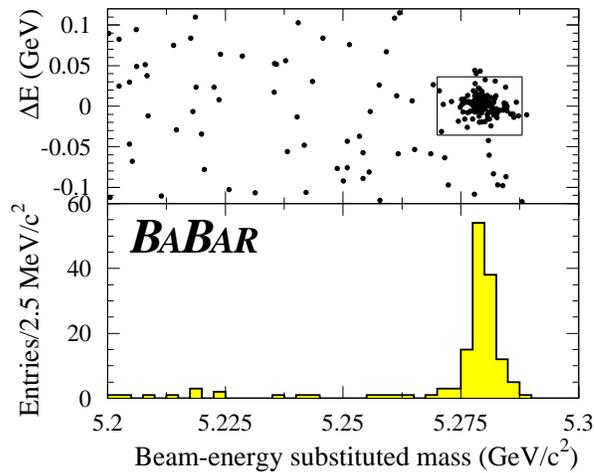,height=3.0in}}
\caption[]{
The signal in the $B^0\to J/\psi K^0_S$, $K^0_S\to \pi^+\pi^-$ mode.
}
\label{fig:cpeventspm}
\end{figure}

\begin{figure}
\centerline{\psfig{figure=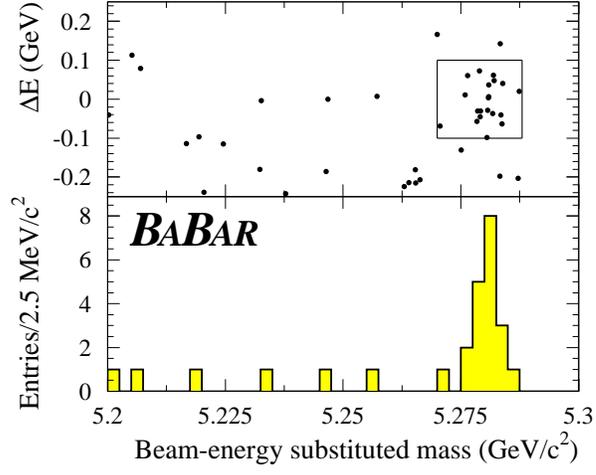,height=3.0in}}
\caption[]{
The signal in the $B^0\to J/\psi K^0_S$, $K^0_S\to \pi^0\pi^0$ mode.
}
\label{fig:cpeventsneutral}
\end{figure}

\begin{figure}
\centerline{\psfig{figure=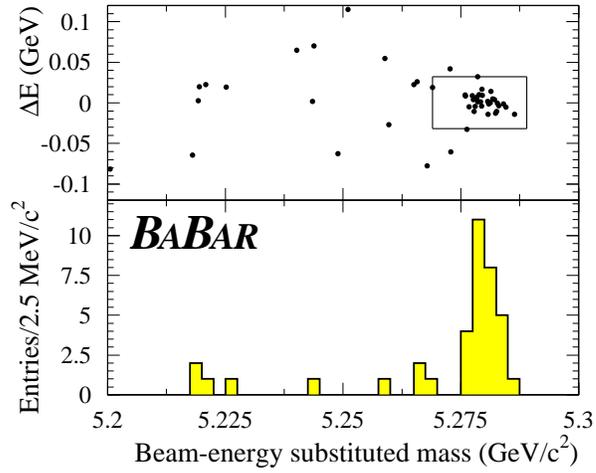,height=3.0in}}
\caption[]{
The signal in the $B^0\to \psi(2S) K^0_S$, $K^0_S\to \pi^+\pi^-$ mode.
}
\label{fig:cpeventspsip}
\end{figure}

 The CP sample consists of events reconstructed in the following modes
\begin{center}
\begin{tabular}{ll}
$B^0\to J/\psi K_S^0$ &  ($K_S^0\to \pi^+\pi^-$), \\
$B^0\to J/\psi K_S^0$ &  ($K_S^0\to \pi^0\pi^0$), \\
$B^0\to \psi(2S)K_S^0$ & ($K_S^0\to \pi^+\pi^-$), \\
\end{tabular}
\end{center}
where the $J/\psi$ and the $\psi(2S)$ are reconstructed in both the $e^+e^-$
and $\mu^+\mu^-$ channels, the $\psi(2S)$ is also reconstructed in the
$J/\psi\pi^+\pi^-$ channel. In the $e^+e^-$ decays, bremsstrahlung photon
recovery is attempted. For more details about the event reconstruction 
see Ref.~\cite{eventreconstruction}. We obtain a total of 168 events in these
modes, as shown in Figures~\ref{fig:cpeventspm}-\ref{fig:cpeventspsip}. 
The yields and purities are
summarized in Table~\ref{tab:cpyield}.

 To measure the performance of the tagging, and in particular to determine
the wrong tag fraction, a sample of exclusively reconstructed $B$ mesons
is used. The modes used and the yields are summarized in 
Table~\ref{tab:brecoyield}. The reconstruction of these samples are 
detailed in Ref.~\cite{lifetime} and \cite{tagmix}.

\subsection{Vertex separation}

 For events where one exclusively reconstructed $B$ meson has been found, the
vertex of the other $B$ decay is determined by trying to combine 
all other tracks in the event.
Candidate tracks that form a good separate vertex, 
e.g., a $K^0_S$, are combined to form a neutral candidate, which is used 
instead of the daughter tracks in the vertex determination. Tracks are
removed if they contribute more than 6 to the $\chi^2$ of the vertex fit.
Events are also rejected if $|\Delta z|>3$ mm
or if $\sigma_{\Delta z}>$ 400 $\mu$m.

 The time resolution function is parametrized as a sum of two Gaussians,
\begin{equation}
{\cal R}(\Delta t;\hat a)= \sum_{i=1}^2{f_i\over \sigma_i\sqrt{2\pi}}
               \exp(-(\Delta t-\delta_i)^2/2\sigma_i^2).
\end{equation}
The resolution parameters, $\sigma_i$, are taken as a scale factor, 
${\cal S}_i$,
times the calculated resolution based on the the tracking errors.
The parameters for the second, wider, Gaussian is fixed from Monte Carlo,
the parameters for the first Gaussian is determined from the combined
fit for mixing and the wrong tag fractions, see Section~\ref{sec:tagmix}.
We also allow for a wide term, $f_w$, with a resolution of 1.8 ps,
and no bias, to handle outliers. The parameters of the resolution
function are summarized 
in Table~\ref{tab:resparm}.

\subsection{Flavor tagging}
\label{sec:tagmix}

\begin{figure}
\centerline{\psfig{figure=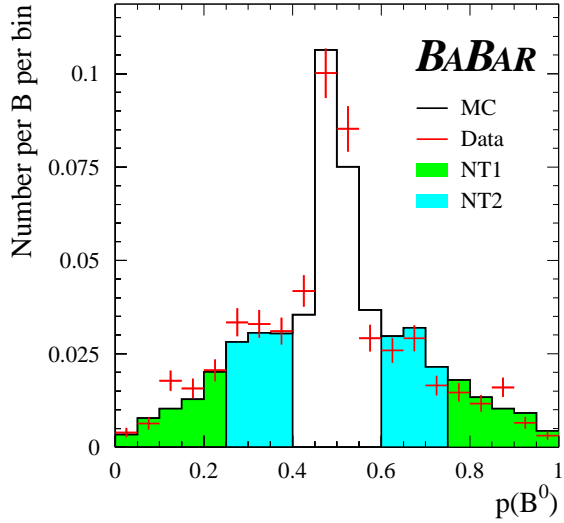,height=3.0in}}
\caption[]{
The output from the neural network for $B$ flavor tagging. The
data is shown as points with error bars and the histogram shows
the Monte Carlo. Events close to 1 (0) are likely to be $B^0$ ($\bar B^0$)
tags. Events near 0.5 provide no taging information and are not used.
}
\label{fig:neuralnetwork}
\end{figure}

 Each event with an exclusively reconstructed $B^0$ decay is assigned a
tag
as a $B^0$ or a $\bar B^0$ if the rest of the event satisfies the 
criteria for one of several tagging categories. These tagging categories
are constructed such that each event will only belong to one category. The
first category uses primary leptons to determine the flavor. If the 
event contains an identified lepton, electron or muon, with center 
of mass momentum greater than 1.1 GeV/c the
event is tagged as a $B^0$ ($\bar B^0$) if the charge of the lepton is 
positive (negative). The second category uses charged particles identified
as Kaons. If the sum
of Kaons charges is positive (negative) the event is assigned a 
$B^0$ ($\bar B^0$) tag. If the lepton tags and Kaon tags disagree no
tag is assigned in these categories. The last two categories, {\tt NT1}
and {\tt NT2}, are assigned based on the output of a neural network. 
The neural network combines information about Kaons, leptons, soft pions and
the stiffest track in the event to form an output that distinguishes 
between $B^0$ and $\bar B^0$ tags. The output from the neural network is shown 
in Figure~\ref{fig:neuralnetwork}. The two tagging categories are
defined such that {\tt NT1} corresponds to the events which has the best 
separation and {\tt NT2} to the events that has slightly worse separation.
The events, in the middle of Figure~\ref{fig:neuralnetwork}, which have very
little separation are not used.

 The figure of merit for each tagging category is the effective 
tagging efficiency, $Q_i=\epsilon_i(1-2w_i)^2=\epsilon_i {\cal D}_i^2$, where
$\epsilon_i$ is the fraction of events assigned to category $i$ and 
$w_i$ is the fraction that had the wrong tag assigned. 

To determine the 
wrong tag fraction we use the sample of exclusively reconstructed hadronic 
$B$ meson decays. These decays tag the flavor of the 
decaying $B$, so by performing 
a combined tagging and mixing analysis we can determine the wrong tag
fraction for each category. For the events in each tagging
category we perform a fit to
\begin{equation}
{\cal H}_{\pm}(\Delta t; \Gamma, \Delta m_{B_d}, {\cal D}, \hat a)=
	{1\over 4}\Gamma e^{-\Gamma |\Delta t|}[1\pm {\cal D}
              \times \cos\Delta m_{B_d} \Delta t]\otimes 
           {\cal R}(\Delta t;\hat a),
\end{equation}
where $+$ are unmixed events and $-$ are mixed events. A log-likelihood
is formed by
\begin{equation}
\ln {\cal L} = \sum_i [ \sum_{\rm unmixed} \ln{\cal H}_{+}
      (\Delta t; \Gamma, \Delta m_{B_d}, {\cal D}_i, \hat a)+
 \sum_{\rm mixed} \ln{\cal H}_{-}(\Delta t; \Gamma, \Delta m_{B_d}, {\cal D}_i, \hat a)]
\end{equation}
where $i$ runs over the tagging categories.
Additional terms are added to the probability density functions 
to describe the contributions from backgrounds, details are
given in~\cite{tagmix}.

 The results of the fit are shown in Table~\ref{tab:tagmix}. The total
tagging efficiency is $76.7\pm0.5$\% with an effective tagging efficiency,
$Q$, of $27.9\pm0.5$\% (statistical errors only).

 When the tagging algorithm is applied to the sample of 168 CP events 
120 events were assigned a flavor tag. Table~\ref{tab:tagging} shows a 
break down per mode and per tagging category of the events in the CP
sample. Of the 120 events tagged, 70 were $B^0$ and 50 were $\bar B^0$
tags. 

\subsection{Systematics}

 Systematic errors were considered from many different sources; input
parameters to the likelihood fit, uncertainties in the time resolution
function and wrong tag fractions. The $B^0$ lifetime was fixed
to the PDG~\cite{pdg} value $\tau_{B^0}=1.548$ ps, $\Delta m_{B_d}$ was
also fixed to the PDG value, $\Delta m_{B_d}=0.472\ \hbar$ps$^{-1}$.
Varying the values of these parameters give the uncertainties on
$\sin2\beta$ for $\tau_{B^0}$ and $\Delta m_{B_d}$ of 0.002 and 0.015
respectively.

 The time resolution function was determined in a high statistics
sample of fully reconstructed $B^0$ events. We vary the parameters
in the time resolution function by 1 statistical standard deviation
and assign a systematic error on $\sin2\beta$ of 0.019. To study
the sensitivity on the bias in $\Delta t$, we allowed the bias of
the second Gaussian to increase to 0.5 ps. This results in a change
of 0.047 on $\sin2\beta$ and is assigned as a systematic uncertainty. 
The sensitivity to this bias is due to the different number of
events that are tagged as $B^0$ and $\bar B^0$.

 The mistag fractions are determined from exclusively reconstructed
$B^0$ and $\bar B^0$ mesons. Sources of systematic uncertainty come
from the presence of backgrounds in these samples and possible 
differences between the tagging performance in the CP sample and the
hadronic samples. The details about the accounting of backgrounds
in the hadronic samples are given in Ref.~\cite{tagmix}. The systematic
uncertainty on $\sin2\beta$ from the measured mistag fractions is
estimated to be 0.053. A rather conservative systematic error of
0.050 on $\sin2\beta$ is assigned for a possible difference between
the tagging performance between the CP sample and the exclusive 
hadronic sample. 

 The CP sample is estimated to contain a background fraction of
$(5\pm3)$\%. Backgrounds from, e.g., $u$, $d$, and $s$ continuum
events contributes primarily at small values of $\Delta t$ and
hence do not contribute much to the determination of $\sin2\beta$.
We estimate that the effective background is 3\% and correct for the
background by increasing the apparent asymmetry by 1.03. A fractional
systematic error of 3\% is assigned on the asymmetry to cover the
uncertainty in the size of the background as well as any possible 
CP asymmetry in the background.

 The systematic errors are summarized in Table~\ref{tab:sys}, and
a total systematic uncertainty of 0.09 on $\sin2\beta$ is obtained.

\subsection{Results}

 The analysis was carried out blind to eliminate any possible
experimenters bias. The blinding technique hid both the 
result of the likelihood fit for $\sin2\beta$, as well as
the CP asymmetry in the $\Delta t$ distribution. The error
on the asymmetry was not hidden. The value of $\sin2\beta$ was hidden
by adding an arbitrary offset and flipping the sign. The CP
asymmetry in the $\Delta t$ distribution was hidden by adding an offset
and, on an event by event basis, multiply $\Delta t$ with the 
sign of the tag.

 This allowed us to carry out many systematic studies while 
keeping the value of $\sin2\beta$ hidden. In particular, the
whole analysis procedure, including event selection, was fixed
prior to unblinding the value of $\sin2\beta$.

 Using Eq.~\ref{eq:sintwobeta}, we perform a likelihood fit to determine one
single parameter, $\sin2\beta$. 
The mistag fractions, $w_i$, and the resolution function parameters,
$\hat a$ are taken from Tables~\ref{tab:tagmix} and~\ref{tab:resparm}
respectively. We obtain 
\begin{equation}
\sin2\beta=0.12\pm0.37\pm0.09.
\end{equation}

\begin{figure}
\centerline{\psfig{figure=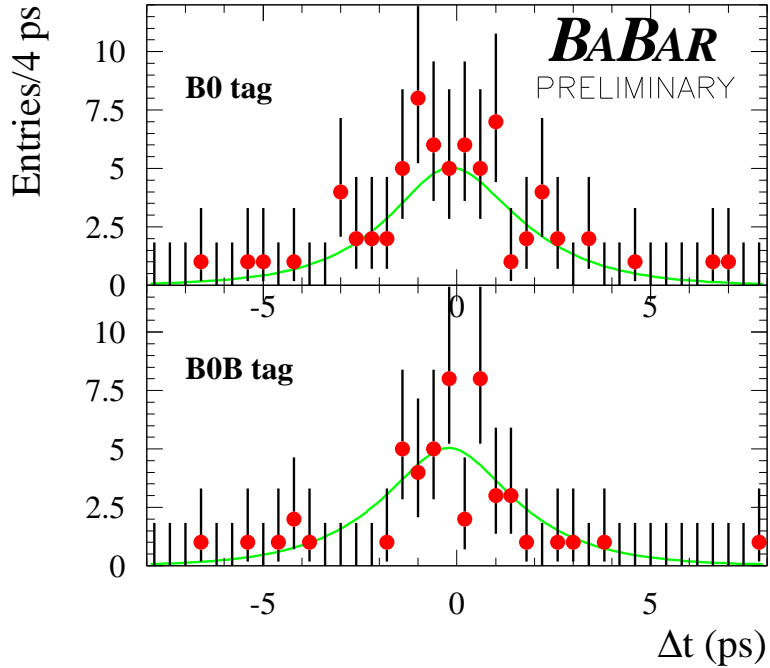,height=4.0in}}
\caption[]{
The $\Delta t$ distributions for events in the CP sample tagged
as  $B^0$ and $\bar B^0$ respectively. The asymmetry in the 
number of $B^0$ and $\bar B^0$ tags are apparent in the plot.
The shift to negative values of $\Delta t$ is due to the small
bias, about 0.2 ps, in the time resolution.
}
\label{fig:timedist}
\end{figure}

\begin{figure}
\centerline{\psfig{figure=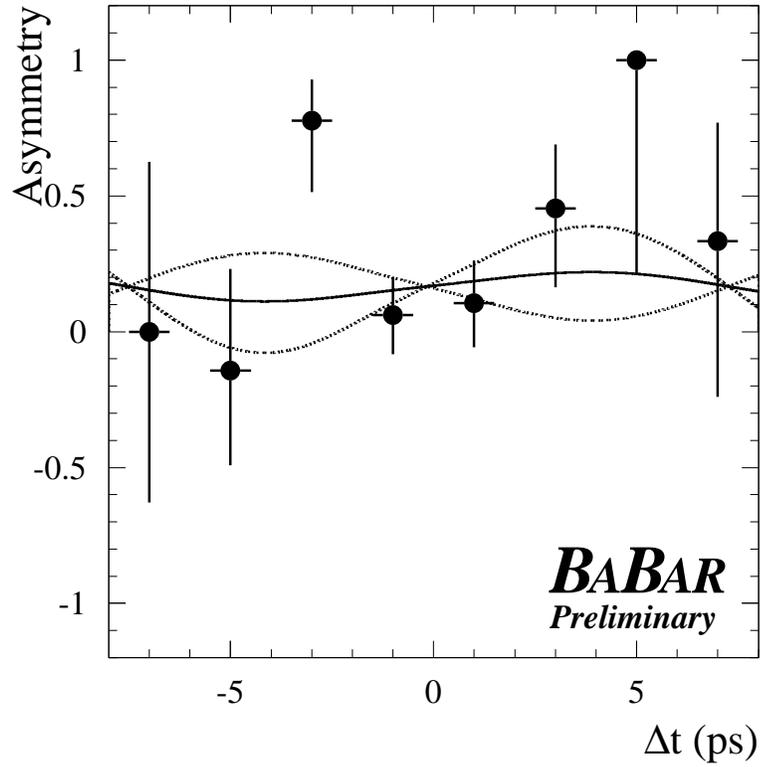,height=4.0in}}
\caption[]{
The raw $B^0-\bar B^0$ asymmetry 
$(N_{B^0}-N_{\bar B^0})/(N_{B^0}+N_{\bar B^0})$, with binomial errors,
as a function of $\Delta t$. The solid line is the asymmetry for our
central value of $\sin2\beta$. The dotted lines represents one
statistical standard deviations from the central value. 
}
\label{fig:asymmetry}
\end{figure}

 Table~\ref{tab:subsample} shows a breakdown of the fit for different
modes and tagging categories. Figure~\ref{fig:timedist} shows the 
$\Delta t$ distributions for $B^0$ and $\bar B^0$ tagged events, the
different yields of $B^0$ and $\bar B^0$ tags is apparent in this plot
as well as in Figure~\ref{fig:asymmetry}, which shows the raw
asymmetry.

 To validate the analysis several cross checks have been made. In 
particular, we have used the charmonium non-CP samples such as
$B^+\to J/\psi K^+$ and the self tagging $B^0$ mode, 
$J/\psi K^{*0}$, $K^{*0}\to K^+\pi^-$, to establish that we do not observe 
any time dependent asymmetry in these modes. We also use the large
samples of exclusively reconstructed charged and neutral $B$ mesons 
in this check. Table~\ref{tab:check} summarizes the fits for an 
apparent CP asymmetry 
in these modes. The result is consistent with no time dependent asymmetry
in these modes.

Other important validation measurements are the 
lifetime and mixing measurements~\cite{lifetime,tagmix}, also
discused in detail in Rainer
Bartoldus contributions to this conference.

\begin{figure}
\centerline{\psfig{figure=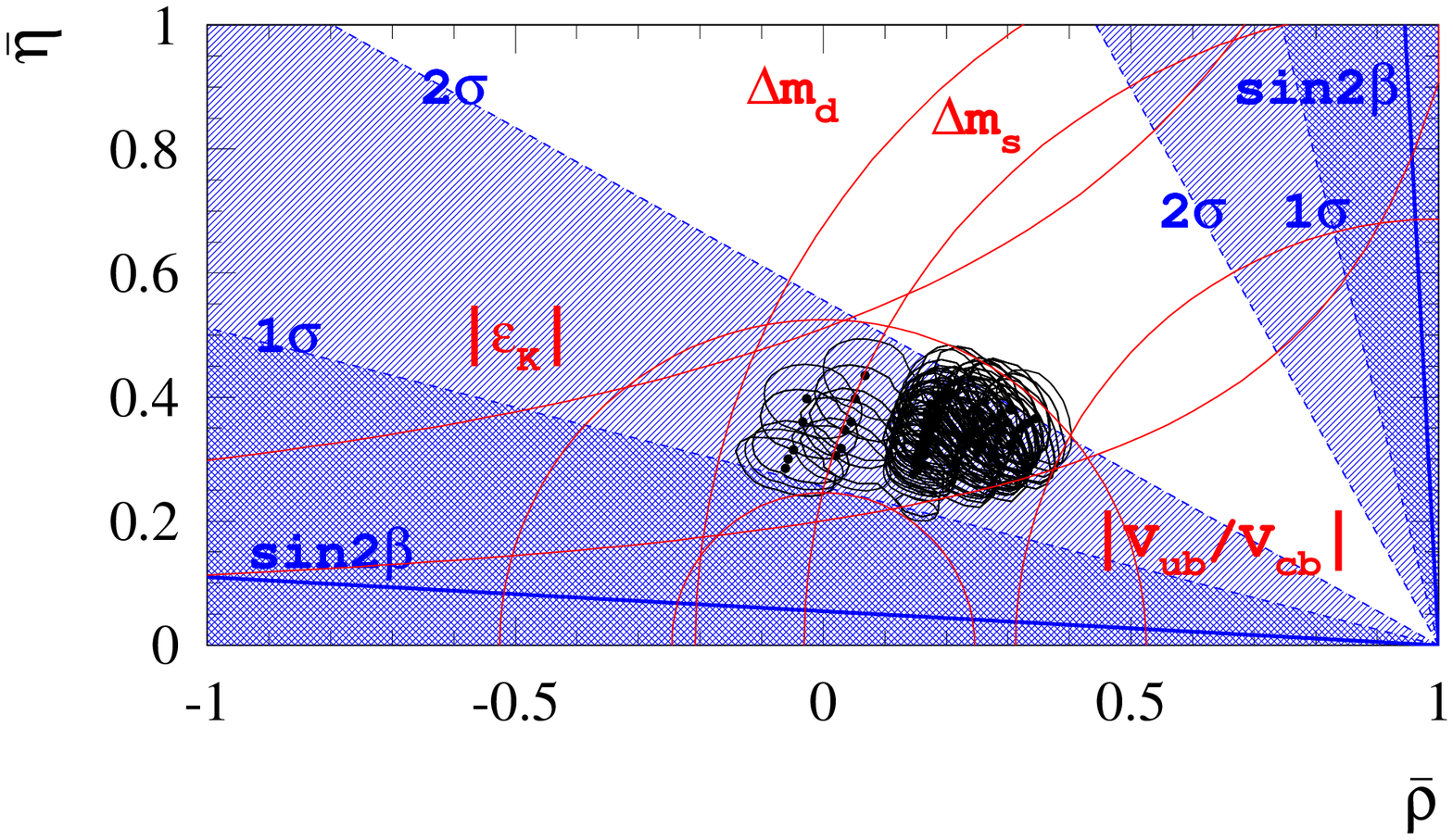,height=4.0in}}
\caption[]{
The constraints in the $\hat\rho-\hat\eta$ planes, and the
result of our determination of $\sin2\beta$ overlayed.
The following measurements are used: 
$\left| V_{cb} \right| = 0.0402\pm0.017$, $ \left| V_{ub}/V_{cb} 
    \right| = \left< \left| V_{ub}/V_{cb} \right| \right> 
    \pm 0.0079$, $\Delta m_{B_d} = 0.472\pm0.017 \, \hbar {\rm ps}^{-1}$ 
    and $\left| \epsilon_K  \right| = \left( 2.271 \pm 0.017 \right) 
    \times 10^{-3}$, and for ${\rm \Delta} m_{B_s}$ the set of 
    amplitudes corresponding to a $95\%$CL limit of $14.6 \, \hbar {\rm ps}^{-1}$.   
We scan the model-dependent parameters $\left< \left| V_{ub}/V_{cb} \right| \right>$, $B_K$,
$f_{B_d} \sqrt{ B_{B_d} } $ and $\xi_s$, in the range $\left[ \, 0.070, \, 0.100 \, 
     \right]$, $\left[ \, 0.720, \, 0.980 \, \right]$, $\left[ \, 185, \, 255 \, 
     \right]$ MeV and $\left[ \, 1.07, \, 1.21 \, \right]$, respectively.   
$\sin2\beta = 0.12\pm0.37{\rm (stat)}$ is represented by cross-hatched 
regions corresponding to 
one and two statistical standard deviations.
}
\label{fig:rhoeta}
\end{figure}

 Figure~\ref{fig:rhoeta} shows the constraints in the $\rho-\eta$
plane and the preliminary BABAR measurement of $\sin2\beta$. The allowed
area in the plot does not include any constraint from our measurement
of $\sin2\beta$.

\section{Conclusion}

 BABAR has reported preliminary results based on a dataset of
9.0 fb$^{-1}$ recorded
at the $\Upsilon(4S)$ resonance from January to July in 2000. 
The analysis used
the modes $B^0\to J/\psi K^0_S$ and $B^0\to \psi(2S) K^0_S$ and studied the
time dependent asymmetry in events where the flavor of the other $B$
meson was determined. The preliminary result
\begin{equation}
\sin2\beta=0.12\pm0.37\pm0.09
\end{equation} 
was obtained. An analysis of the full 2000 dataset is now in progress,
other modes, including $B^0\to J/\psi K^0_L$, is to be included
to further improve the precision of the measurement of $\sin2\beta$.
Data collected over the next few years should significantly improve the 
precision of this measurement. BABAR and PEP-II will start running again
in February of 2001.

\newpage


\begin{table}[htb]
\begin{center}
\begin{tabular}{|l|c|c|} \hline
Final state  & Yield & Purity (\%) \\ \hline \hline
$J/\psi K_S$ ($K_S \to \pi^+\pi^-$)              &  121 &  96 \\
$J/\psi K_S$ ($K_S \to \pi^0 \pi^0$)             &   19 & 91 \\
$\psi(2S) K_S$ ($K_S \to \pi^+\pi^-$)            &   28 &  93 \\ 
\hline
\end{tabular}
\caption{The yield of events in the different modes for the CP sample.
}
\label{tab:cpyield}
\end{center}
\end{table}

\begin{table}[htb]
\begin{center}
\begin{tabular}{|l|c|c|} \hline
Final State & Yield & Purity (\%)       \\ \hline \hline
$D^{*-}\pi^+$    & $622\pm 27$   &   90   \\
$D^{*-}\rho^+$   & $419\pm 25$   &   84   \\
$D^{*-}a_1^+$    & $239\pm 19$   &   79   \\
$D^{-}\pi^+$    & $630\pm 26$   &   90   \\
$D^{-}\rho^+$   & $315\pm 20$   &   84   \\
$D^{-}a_1^+$    & $225\pm 20$   &   74   \\ \hline
Total   & $2438\pm 57$   &  85   \\ \hline\hline
$\bar D^{0}\pi^+$ & $1755\pm 47$   &   88   \\
$\bar D^{*0}\pi^+$ & $543\pm 27$   &   89   \\ \hline
Total         & $2293\pm 54$   &   88   \\ \hline \hline
$D^{*-}\ell^+\nu$ & $7517\pm 104$   &   84   \\ \hline
\hline
\end{tabular}
\caption{The yield of events in the different modes for $B$ mesons
reconstructed in hadronic and semileptonic modes.
}
\label{tab:brecoyield}
\end{center}
\end{table}

\begin{table}[htb]
\begin{center}
\begin{tabular}{|l|c|c|c|} \hline
Tagging Category & $\varepsilon$ (\%) & $w$ (\%) & $Q$ (\%)       \\ \hline \hline
{\tt Lepton}     & $11.2\pm0.5$ & $9.6\pm1.7\pm1.3$   &  $7.3\pm0.3$  \\
{\tt Kaon}       & $36.7\pm0.9$ & $19.7\pm1.3\pm1.1$  &  $13.5\pm0.3$  \\
{\tt NT1}        & $11.7\pm0.5$ & $16.7\pm2.2\pm2.0$  &  $5.2\pm0.2$  \\
{\tt NT2}        & $16.6\pm0.6$ & $33.1\pm2.1\pm2.1$  &  $1.9\pm0.1$  \\  \hline \hline
all              & $76.7\pm0.5$ &                     &  $27.9\pm0.5$ \\ 
\hline
\end{tabular}
\caption{
The efficiency and wrong tag fractions in the different tagging
categories as determined from the combined tagging and mixing fit to
the hadronic and semileptonic event sample.
}
\label{tab:tagmix}
\end{center}
\end{table}

\begin{table}[htb]
\begin{center}
\begin{tabular}{|l||c|c|c||c|c|c||c|c|c||c|c|c|} \hline
  & \multicolumn{6}{|c||}{\rule[-1pt]{0mm}{14pt}$J/\psi K_S$} 
  & \multicolumn{3}{|c||}{$\psi(2S) K_S$} 
  & \multicolumn{3}{|c|}{$CP$\ sample } \\ \cline{2-13}
 
  &  \multicolumn{3}{|c||} {\rule[-1pt]{0mm}{14pt}($K_S \to \pi^+\pi^-$)} 
  &  \multicolumn{3}{|c||} {($K_S \to \pi^0\pi^0$)} 
  &  \multicolumn{3}{|c||} {($K_S \to \pi^+\pi^-$)} 
  &  \multicolumn{3}{|c|}  {(tagged)} 
  \\ \cline{2-13}
     &  $B^0$\ & \rule[-1pt]{0mm}{14pt}$\bar B^0$\ & all &  $B^0$\ & 
       $\bar B^0$\ & all &  $B^0$\ & $\bar B^0$\ & all &  $B^0$\ &
 $\bar B^0$\ & all \\ \hline \hline 
{\tt Electron}   
  & 1  &  3 & 4  
  & 1  &  0 & 1  
  & 1  &  2 & 3 
  & 3  &  5 & 8      \\
{\tt Muon}       
  & 1  &  3 & 4  
  & 0  &  0 & 0  
  & 2  &  0 & 2 
  & 3  &  3 & 6      \\
{\tt Kaon}       
  & 29 & 18 & 47 
  & 2  & 2  & 4  
  & 5  & 7  &12 
  & 36 & 27 & 63     \\
{\tt NT1}       
  &  9 &  2 & 11 
  & 1  & 0  & 1  
  & 2  & 0  & 2 
  & 12 & 2  & 14     \\
{\tt NT2}       
  & 10 &  9 & 19 
  & 3  & 3  & 6  
  & 3  & 1  & 4 
  & 16 & 13 &  29    \\
\hline \hline
{\tt Total}      
  &  50& 35 & 85 
  & 7  & 5  &12 
  &13  &10  &23  
  & {70} & {50} & {120 }   \\

\hline
\end{tabular}
\caption{
The result of tagging applied to the CP sample and broken down
by tagging category and mode.
}
\label{tab:tagging}
\end{center}
\end{table}

\begin{table}[htb]
\begin{center}
\begin{tabular}{|cc|cl|} \hline
   \multicolumn{2}{|c|}{Parameter} & \multicolumn{2}{c|} {Value}    \\ \hline \hline
 $\delta_1$  & (ps)    & $-0.20\pm0.06$  & from fit     \\
 ${\cal S}_1$&   & $1.33\pm0.14$       & from fit     \\
 $f_{w}$       & (\%)  & $1.6\pm0.6$     & from fit     \\
 $f_2$       & (\%)  & $25$              & fixed        \\
 $\delta_2$  & (ps)  & $0$             & fixed        \\
 ${\cal S}_2$ &  & $2.1$               & fixed        \\
\hline
\end{tabular}
\caption{The parameters of the resolution function as determined from the 
sample of fully reconstructed hadronic $B^0$ candidates.
}
\label{tab:resparm}
\end{center}
\end{table}

\begin{table}[htb]
\begin{center}
\begin{tabular}{|l|c|} \hline
 sample                                    &  $\sin2\beta$  \\ \hline \hline
 CP sample                               &  {\bf 0.12}$\pm${\bf 0.37}  \\  
\hline
 \ \ $J/\psi K_S$ ($K_S \to \pi^+ \pi^-$) events  &  $-0.10 \pm 0.42$   \\  
 \ \ other CP events                           &  $0.87 \pm 0.81$   \\  
\hline
 \ \ {\tt Lepton}                         &  $1.6 \pm 1.0  $  \\
 \ \ {\tt Kaon}                           &  $0.14\pm 0.47   $   \\
 \ \ {\tt NT1}                            &  $-0.59\pm0.87  $   \\
 \ \ {\tt NT2}                            &  $-0.96\pm 1.30  $   \\
\hline 
\end{tabular}
\caption{The result of the fit for $\sin2\beta$ broken down by
event and tagging category.
}
\label{tab:subsample}
\end{center}
\end{table}

\begin{table}[htb]
\begin{center}
\begin{tabular}{|l|c|} \hline
Systematic Error & Uncertainty on $\sin2 \beta$       \\ \hline \hline
$\tau^0_B$               & $0.002$ \\
$\Delta m_d$             & $0.015$ \\
$\Delta z$ resolution    & $0.019$ \\
time resolution bias     & $0.047$ \\
measured mistag fraction & $0.053$ \\
$CP$ versus non-$CP$ sample & \\
        mistag fraction & $0.050$ \\
$B^0$ versus $\bar B^0$ mistag fraction     & $0.005$ \\
background in $CP$ sample & $0.015$ \\ \hline
total systematic error   & $0.091$ \\ \hline \hline
\end{tabular}
\caption{
Summary of sources of systematic errors in the determination of $\sin2\beta$.
}
\label{tab:sys}
\end{center}
\end{table}

\begin{table}[htb]
\begin{center}
\begin{tabular}{|l|c|} \hline
 Sample                                    & Apparent CP asymmetry  \\ 
\hline \hline
 Hadronic charged $B$ decays                  &   $0.03 \pm 0.07$   \\  
\hline
 Hadronic neutral $B$ decays            &   $-0.01 \pm 0.08$  \\  
\hline
 $J/\psi K^+$                               &   $0.13\pm 0.14$    \\ 
\hline 
 $J/\psi K^{*0}$ ($K^{*0} \to K^+ \pi^-$)   &  $ 0.49 \pm 0.26$   \\ 
\hline 
\end{tabular}
\caption{Results of fitting for apparent CP asymmetries in various 
charged or neutral flavor tagging $B$ samples. 
}
\label{tab:check}
\end{center}
\end{table}

\newpage

\begin{thebibliography}{99}

\bibitem{babarphysbook}
P. H. Harrison and H. R. Quinn, eds., ``The BABAR Physics Book'', SLAC-R-504 
(1998) and references therein.

\bibitem{babarpaper}
BABAR Collaboration, B. Aubert {\it et al.}, ``The first year of the BABAR
experiment at PEP-II'', BABAR-CONF-00/17, submitted to the XXX$^{th}$ 
International Conference on High Energy Physics, Osaka, Japan.

\bibitem{eventreconstruction}
BABAR Collaboration, B. Aubert {\it et al.}, ``Exclusive $B$ decays to 
charmonium final states'', BABAR-CONF-00/05, submitted to the XXX$^{th}$ 
International Conference on High Energy Physics, Osaka, Japan.

\bibitem{pdg}
Particle Data Group, D. E. Groom {\it et al.}, 
Eur. Phys. Jour. C {\bf 15}, 1 (2000)

\bibitem{lifetime}
BABAR Collaboration, B. Aubert {\it et al.}, ``A measurement of
the charged and neutral $B$ meson lifetimes using fully reconstructed 
decays'', 
BABAR-CONF-00/07, submitted to the XXX$^{th}$ 
International Conference on High Energy Physics, Osaka, Japan.



\bibitem{tagmix}
BABAR Collaboration, B. Aubert {\it et al.}, ``A measurement of the  
$B^0\bar B^0$ oscillation frequency and determination of flavor-tagging 
efficiency using semileptonic and hadronic $B$ decays'', 
BABAR-CONF-00/08, submitted to the XXX$^{th}$ 
International Conference on High Energy Physics, Osaka, Japan.



\end{thebibliography}
\end{document}